\documentclass[a4paper,12pt]{article}
\usepackage{cite}
\usepackage{amssymb,amsmath}
\newcommand{\be}{\begin{equation}}
\newcommand{\ee}{\end{equation}}
\newcommand{\bea}{\begin{eqnarray}}

\newcommand{\eea}{\end{eqnarray}}

\begin{document}

\title{A Comment or two on Holographic Dark Energy}

\author{A.J.M. Medved \\ \\
Physics Department \\
University of Seoul \\
 Seoul 130-743\\
Korea \\
E-Mail(1): allan@physics.uos.ac.kr \\ 
E-Mail(2):  joey\_medved@yahoo.com \\ \\}

\maketitle
\begin{abstract}

It has, quite recently, become fashionable to study a
certain class of 
holographic-inspired models for the dark energy. These
investigations
have, indeed, managed to make some significant
advances towards explaining
the empirical data. Nonetheless, surprisingly little
thought has been given
to conceptual issues such as the composition and the
very nature of the 
implicated energy source. In the current discourse, we
attempt to fill
this gap by the way of some speculative yet logically
self-consistent
arguments. Our construction takes us along a path that
begins with an 
entanglement entropy and ends up at a Hubble-sized gas
of exotic particles. 
Moreover, our interpretation of the dark energy turns
out to be 
suggestive of a natural resolution to the
cosmic-coincidence problem.

\end{abstract}
\newpage
\section{Background and Buildup}

It is well acknowledged that our present-day universe
is
 (or at least appears to be) 
 in a phase of  cosmological acceleration
\cite{acc-uni}. The simplest
 and
(perhaps)  most aesthetically pleasing
explanation for this phenomenon would be a
cosmological constant.
(As was originally proposed by Einstein, albeit
with a much different motivation in mind
\cite{ein-cc}.) On the
basis of Occam's Razor, this might well be the end of
the story, except
for a few  points of notable infamy \cite{weinberg}:

 {\it (i)}  The
 {\it cosmological-constant problem} (version 1), or
why is the 
empirically  based  value 
of the cosmological constant so small in comparison to
the Planck scale
 (which is, as prescribed by quantum field theory,
 the natural scale that one would associate with the
energy of the vacuum)? For future reference, the
discrepancy between
the two scales (empirical and Planck) is   the
staggering
amount of at least 120 orders of magnitude.   \\
{\it (ii)} The cosmological-constant problem (version
2), 
or given that the cosmological constant 
is so small, why is it not simply zero? To rephrase,
suppose there is a
fundamental  symmetry at play (along the lines of
supersymmetry) that is indeed responsible for  
Planck-scale
 cancellations  
in the vacuum energy. Then what mechanism ends up
breaking this 
symmetry at such an unnaturally small
number ($10^{-120}$)? \\
{\it (iii)} The {\it cosmic-coincidence problem}, or
why do we happen 
to live in the
 (cosmologically speaking) briefest of  eras when the
energy density of
 matter
and the cosmological constant are virtually the same
({\it i.e.}, 
within an order
of magnitude)?

In view of a cosmological {\it constant} ({\it per
se}) being 
somewhat problematic,
many proposals have been suggested that 
supplant this fixed quantity with a more
dynamical source. By current conventions, 
any such alternative (as well as the constant itself)
falls under the generic classification of being a 
{\it dark energy}.~\footnote{For further discussion on
the
cosmological-constant problems, various proposals 
for the dark energy, {\it etcetera},
see the reviews in \cite{cc-rev1,cc-rev2,cc-rev3}.} 
For instance, the potential 
energy for a slowly varying scalar field  has 
often been 
nominated  for the role of a dark energy; notably, 
such (so-called) {\it quintessence} models 
 arise quite naturally out of the framework of string
theory
\cite{quin}
and are  analogous to the  {\it inflaton}-description
of inflation \cite{infl,infl2}.
[The latter being the  (generally accepted) earlier
era of 
cosmic acceleration.]

There are, of course, a  litany of
dark energy proposals in the literature. Here, we will
focus on 
what has become commonly known as the {\it
holographic} model of dark 
energy. (See \cite{m-li} for one of the seminal
renditions,
\cite{N-myung} for further discussion and 
\cite{hde-rev} for an overview with almost too many
references.)
The basic premise is inspired by the  famed 
{\it holographic principle}
\cite{bekenstein,thooft,susskind,bousso},
with particular emphasis towards a pertinent
observation made in 
\cite{cohen}. We will now proceed to elaborate
on this circle of ideas.

In spite of its  many manifestations in the
literature, the 
essence of the holographic principle is 
remarkably simple: a black hole represents, in a very 
fundamental sense, the maximally entropic
object  for a fixed amount of energy. As applied to a 
given (quantum)
 field theory, the principle advocates
$S_{QFT} < S_{BH}$ or, as long as  our field theory
comes equipped
 with
an ultraviolet-energy cutoff $\Lambda$ and an
infrared-length cutoff
 $L$,
it would follow that
\be
L^3 \Lambda ^3 \;\lesssim \; L^2 M_{P}^2 \;.
\label{1}
\ee
Here, we have only used the extensivity of the  
(unconstrained) field-theoretic entropy,
 the area--entropy law for black holes \cite{area},
and the (near)
 saturation
of the initial inequality at the respective
cutoffs.~\footnote{Further
 note
that we choose, throughout, to   work with a 
four-dimensional
  spacetime,
 set all fundamental constants --- except for the
Planck mass $M_P$ ---
to unity, and consistently ignore all inconsequential
numerical
 factors.} 

By very similar reasoning, one can talk about
constraining the energy.
In fact, by virtue of the {\it hoop conjecture}
\cite{hoop} (or,
 equivalently,
the Schwarzschild limit),
 the field-theoretic energy
should be limited by the  linear size of the system.
Then extensivity
of the former and saturation at the cutoffs leads to
\be
L^3 \Lambda ^4 \;\lesssim \; L M_{P}^2 \;.
\label{2}
\ee
Since $L$ is, by hypothesis, the longest (meaningful)
length scale in
 the 
theory, it follows that this second relation is the
one that
maximally constrains $\Lambda$. We are thus obliged to
use
the latter in obtaining the maximal energy density.
Consequently,
\be
\rho \;\lesssim \; \Lambda^4 \; \lesssim \;
{M_P^2\over L^2}\; .
\label{3}
\ee

It is this  very last relation that served as the
impetus for the
 notion
of a ``holographic dark energy". Namely, it has been
suggested
that the form of the dark energy should conform
precisely
with
\be
\rho_{DE} \; \leq \; c^2{M_P^2\over L^2} \;,
\label{4}
\ee
where $c^2$ is a numerical factor 
which is always taken to be of the order of unity
and  can  typically be constrained by the way of
observational
data. Predominantly (but  not universally
\cite{non-satur}),
this has been  regarded as a literal equality; meaning
that the
 saturation
point is to be chosen {\it a priori}. 

What has never been particularly clear is to what
choice should be made
for the infrared cutoff $L$. The usual
suspects are the   apparent horizon or Hubble radius
$H^{-1} = a/{\dot a}\;$,~\footnote{Here, $a=a(t)$ is
the cosmological
scale factor and a dot denotes a differentiation with
respect
to the cosmic time $t\;$.  We will be presuming a flat
universe in
our discussions (hence, the equivalency
of the apparent horizon and the Hubble radius),
 although almost nothing that is said really
depends upon this distinction.}
the future (event) horizon
and the past (particle) horizon.~\footnote{This short
list
is by no means meant to be exhaustive. For instance,
there is 
the ``generalized" model of dark energy (for which
$L$ is permitted to depend, functionally, on all three
of the above) \cite{N-od} and the
``agegraphic" model  of dark energy (whereby the
age of the universe determines $L$) \cite{cai}.}
Respectively, the latter two  are as follows:
\be
d_{f} \;=\; a\int^{\infty}_t{dt\over a} \;,
\label{5}
\ee
\be
d_{p} \;=\; a\int^{t}_0{dt\over a} \;.
\label{6}
\ee

At  a  first glance of the literature, it would appear
that   
the future horizon is empirically favored
\cite{m-li} and the apparent horizon is
emphatically  ruled out \cite{hsu}. Nevertheless, all
bets remain on
once the model is suitably complicated to include 
spatial curvature \cite{m-li2} and/or
interactions
with the matter sector \cite{time-grav,interact}, a
time-varying
 gravitational
coupling \cite{time-grav}, the aforementioned
non-saturated
(holographic) bound \cite{non-satur,N-rand}, or whatever the
flavor of the day
 happens
to be.~\footnote{Again, the reader should consult 
\cite{hde-rev} for 
many other relevant citations.} 

Irrespective of the exact choice for $L$, it is clear
that 
the holographic model of  dark energy has some
attractive features.
Firstly,  it provides, by construction, 
an immediate (albeit somewhat {\it ad hoc})
solution to the cosmological-constant problem of the
present era.
Secondly, the fact that $L$ is (presumably) a
dynamical quantity allows
 us
to hope for even more:  A fully realistic form of the
model might be
 able
to account for the dark energy in all 
other relevant  eras (particularly, inflation)
and accommodate a decelerating to 
accelerating phase (and {\it vice versa})
at the appropriate junctures.
Thirdly, the holographic model has had, 
when suitably generalized (see the paragraph above),
some degree of success in  resolving even  the
cosmic-coincidence
 problem.

Nonetheless, there are (at least) a few aspects of the
overall
 framework that
remain puzzling. Clearly, this dark energy model 
is somewhat contrived; having been
posed in such a manner so as to defer, rather than
resolve, the 
very issues that inspired it.
And let us ponder the following: What is the substance
(for lack
of a better word)  that constitutes or underlies 
the implied dark energy? If the answer
is some new type of particle, then where does this new
entity fit into
the grander scheme of things? (For instance,
is it agreeable with the standard model?) If the
answer
is some manifestation of  ``pure geometry", 
 then how is the use of the
field-theoretic bounds to be justified in the first
place?
If the answer is none of the above, then are we 
really
better off here than with a fixed cosmological
constant?

Anyways,  justifying and explaining the exact nature
of the dark
energy may be the least of our (conceptual) worries.
The very definition of the holographic dark energy
seems to hinge
on the existence of a physically meaningful infrared
cutoff; which is
 to
say, a physically viable cosmological horizon. On the
other hand,
it is not at all evident that one can have such a
horizon in the
 absence
of some sort of dark energy. For instance, let us
consider the
popular choice of a future horizon. Without a dark
energy
or a cosmological constant to induce acceleration (and
assuming no other forms of exotic  matter  are
present), a simple
 calculation
reveals that the future horizon is necessarily
infinite.~\footnote{In
short, an accelerating universe and, thereby, finite
future
horizon requires an equation-of-state  parameter
({\it i.e.}, a ratio of pressure to energy density)
 of $\omega<-1/3\;$. [Alternatively, one can use 
the standard Friedmann result of $a\sim
t^{2/3(1+\omega)}$
along with Eq.(\ref{5}).]
If the choices are limited to dust matter
($\omega=0$), radiative
matter ($\omega=1/3$) and spatial curvature
($\omega=-1/3$),
it becomes quite evident that such an inequality is
strictly
 unattainable.
(For future reference, a ``true" cosmological constant
has $\omega=-1$
and the observational bound is $\omega < -0.75$
\cite{acc-uni}.)
$\label{fn}$}
 Hence, there can be no such infrared 
cutoff without first having a dark energy 
and, likewise, there will be no  holographic dark
energy without 
having a cutoff to define it.
This leaves us with a conceptual paradox that is
tantamount
to the proverbial question about ``the chicken or the
egg"!

Admittedly, it would be difficult to explain away  any
of these
points  until there is an established model of quantum
gravity to call
 upon.
So one might argue that any such issues should be
deferred until
that ``messianic  time" is  finally upon us.
But it might, just as well,  be folly to continue
tweaking a model that
 has
no physically motivated rationalization other than
appealing to
a proposed principle of a yet unfounded theory.

On this last note, our current aim is to fill in the
gap (somewhat)
 with
a (n admittedly) speculative proposal for 
the nature of the holographic dark energy.
Our starting point  will be that the dark energy is
induced from an
entropy of entanglement,
and the discussion will be advanced, sequentially, 
from there.
As a way of motivation,  an association
between  dark energy and  entanglement
 appears to be a very natural one in a holographic
context.
To elucidate, an  entanglement entropy has been
proposed   \cite{bh-ent1,bh-ent2} 
as a viable explanation for the
(holographic-inspiring)
black hole area--entropy law.
Moreover, it appears to have an important 
significance \cite{ads-ent1,ads-ent2}  in the most 
rigorous realization of the holographic principle to
date; 
namely, the {\it AdS--CFT correspondence}
\cite{maldacena}. 
[Note that the idea of connecting  the dark energy 
to a quantum  entanglement  is  not meant to be
original: This 
connection was first proposed (concretely)  in
\cite{ent-de} and  then advanced in
\cite{ent-de2,horvat}. 
Although there will inevitably be some overlap between
the entanglement  part of
our discussion and these other works, we feel that the
current 
perspective  and interpretations are clearly 
distinct.~\footnote{And,  for the interested reader,
see \cite{N-greek} for an 
altogether different viewpoint on the conceptual
origin of
the holographic dark energy.}]

The succession of points to be made will be 
organized  as in the following summary: \\
({\bf 1}) Any relevant observer~\footnote{Whenever we
refer to
an ``observer",  it is meant  to be in the most 
passive of senses. For instance,
an observer might simply be  a test particle that is
being used to map
 out a 
specified world line in the spacetime. It should  
not  (necessarily) be interpreted as an
intelligent life form; even if our choice of 
semantics does imply this at times.} 
 will be causally separated from a significant portion
 of the
 universe.
So, for physics to locally make sense, it is natural
(if not absolutely necessary) 
 to trace over these inaccessible
degrees of freedom; thus giving  rise to an  entropy
of
{\it entanglement}. \\
({\bf 2}) By some (yet-to-be-specified)  process,
the entanglement-induced degrees of freedom should be 
``uplifted" to
the status of actual physical particles
 (insofar as the observer in question is concerned).
Although unspecified, this process of {\it uplifting} 
can still be anticipated because of an analogy  
with the Unruh effect \cite{unruh}. \\
({\bf 3}) Given the intuitively  
expected (and also substantiated
\cite{brustein2,brustein3}) 
form for an entanglement energy,
it can (and will) be demonstrated 
that the  {\it energetics} of the induced particles is
in complete compliance with Eq.(\ref{4}) 
for the (holographic) dark energy density. \\
({\bf 4}) Although  situated (at least in a
holographic sense) 
at a causally defined horizon,
the induced matter  will be shown to be inherently
{\it non-localized}; necessarily filling up the
horizon interior
in the guise of an inert gas of long-wavelength
particles. In this way,
 it can be said that a space-filling dark energy has
truly been
 achieved.

We  will now proceed to elaborate on each of these 
points, in turn, and then conclude
with an  overview. \\

\section{The Main Points}

\subsection{Entanglement}

Thanks to the aftermath of inflation, the universe
must be
a vast, vast place; containing, 
at the very least, $e^{60}$ Hubble-sized spheres
(and it could be substantially larger than this lower
bound
 \cite{infl,infl2}).
Meaning that, irrespective of the existence  of a
finite event 
(or some other formal)
 horizon,  an observer
will certainly be out of causal contact with a large
fraction of the
 universe.
[Given that inflation did occur, this must always be
true except at 
very early (pre-inflationary) times and
(depending on the overall energetics)
asymptotically late times. Neither of these eras is of
relevance
to the current discussion.]
The ``rule-of-thumb" estimate for this ``causal
boundary" is
a spherical surface of radius
$H^{-1}$, but the exact location need not concern us
at this 
stage.~\footnote{For a best guess at a rigorous
placement,
we would suggest the ``causal connection scale" as
proposed in
\cite{brustein1}.}
The important point is that, for all practical
purposes,
our observer's universe comes to an abrupt end at a
distance
of (roughly) one Hubble radius away.

If this discussion was limited to strictly 
classical physics, then this could well be the
end of the story. But quantum considerations make for
things to be 
significantly more interesting. Irregardless of the
bounds of
classical causality,
it is safe to assume that most any matter or energy
source
that is outside of this casual  barrier
shares quantum correlations --- or entanglements ---
 with that found inside. (This follows
from ``EPR-like" quantum non-locality \cite{epr} and
the realization
that the {\it entire} universe was once a
Planckian-sized place.) 
Following  the standard ideas of 
quantum measurement \cite{adami}, this means that, to
have 
a sensible quantum description of the boundary
interior,
it is first necessary to trace over the degrees of
freedom in the
 exterior.

More explicitly, let $\vert\Psi>$ represent the
universal wavefunction,
which is  to be 
regarded as a pure (zero-entropy) state. 
Also, let $\vert\phi_A>$ collectively denote a
complete
set of orthonormal states for the 
interior region and let $\vert\phi_B>$ do likewise for
the exterior.
It is known that one can always  write (for a suitable
choice of
complex coefficients $C_{AB}$)
\be
\vert\Psi> \;=\;\sum_A\sum_B C_{AB}
\vert\phi_A>\vert\phi_B>\;.
\label{7}
\ee
Now, given the necessity to trace, 
the appropriate density matrix for an observer in the
interior has the following {\it reduced} form:  
\bea
\rho_A\;&=&\;{\rm Tr_B}\left[ \vert\Psi><\Psi\vert
\right]
\nonumber
\\
& = & \; \sum_B\sum_A\sum_{A^{\prime}}
C_{AB}C^{*}_{A^{\prime}B}
\vert\phi_A><\phi_{A^{\prime}}\vert\;. 
\label{8}
\eea

Let us recall that a gain in entropy can  generally be
associated with
 a loss
in information.
So that, on account of  tracing out the exterior (an
inherently
information-negating process),  
the interior observer  will naturally  assign 
an entropy to her reduced subsystem. 
(Put differently, even though the initial state was
pure,
the reduced density matrix
describes what is now a mixed state.) As should be
well known,
this entropy can be quantified with the use of
the von Neumann formula 
\be
S_A \; =\; -{\rm Tr_A}\left[\rho_A {\rm
ln}\rho_A\right]\;.
\label{9}
\ee 
As long as there are any quantum correlations at all
between the two
subsystems ({\it i.e.}, as long as  $\vert\Psi>$ can
not be
written as a  direct product of  the subsystems), 
 it should be clear that this  
{\it entanglement} entropy will be  a strictly
positive 
quantity~\footnote{To see this, notice that $\rho_A$
is diagonal
and, since all of the states  have been properly
normalized,
each matrix  entry 
 is a non-negative number that does not exceed unity.
The existence of correlations will mean at least
one of these entries is non-vanishing.}
---  in spite of having started off with what was a
pure state of
(necessarily) vanishing entropy.

 Let us next
consider  $S_B\;$; that is,  
the analogous entropy as would be assigned by an
exterior observer
after tracing over the interior states. 
After some minor manipulations, it should not  
be too difficult to convince oneself
that $S_B  =  S_A\;$.
This is, in actuality, generally true: Starting with a
pure state
in a region $V$, which is then split 
up into two   subregions --- say $V_A$ and $V_B$ ---
one finds (after tracing) that this entropic 
equality is always preserved. The point is that this
subdivision can
be arbitrarily disproportionate; 
for instance, taking $V_B \gg V_A\;$, one  would 
still
find an equality between the corresponding entropies.
(In fact, this is
 just
the situation we would expect for the 
division between the boundary interior $A$ and 
exterior $B$.) This oddity would leave one to believe
that the
 entanglement
entropy can only depend on properties that are common
to both
 subsystems. It
would  logically follow that, to leading order, the
entanglement
 entropy should
be proportional to the area of the  
surface that forms the {\it common} boundary between
the
regions.  Reassuringly,  this outcome has indeed been
realized in
  previous
(rigorously done) calculations \cite{bh-ent1,bh-ent2}.

[As it now stands, the above line of reasoning is
somewhat misleading.
The existence of long-range correlations means that,
generally speaking,
the (leading-order) entanglement entropy is expected
to scale with the volume
of the total region \cite{N-hsu} rather than just the
(common) boundary area. 
As for the rigorous  calculations, these 
apply to the special case of the quantum field theory
being in its
ground state. It, therefore, becomes prudent to ask if
the ground state is
a reasonable expectation in the current context.
We would argue ``yes!" on the premise that this
particular entanglement is, in essence, a
quantum-gravitational process; and so the field theory
should inherit an
energy gap
of (presumably) the order of $M_P$. Such a large gap
would then  act to suppress any excitations of the
system out of its ground
state  and,  thus, achieve the desired area scaling
\cite{N-req}. Alternatively, it has been shown 
\cite{N-hsu} that the (holographic) energy
bound of Eq.(\ref{2}) is also a  sufficient condition 
for
the entanglement entropy to scale with the boundary
area.  
Since we are, after all, talking about  
a {\it holographic} dark energy, it seems quite
sensible  that
such a  bound  should (if necessary) be allowed  to
enter directly into the formalism 
as a {\it principle} of the fundamental theory.]

 For our observer's Hubble-sized sphere in
particular, it can now be safely anticipated that 
\be
S_{E}\;\equiv\; S_{A \;{\rm or}\; B} \; \sim \;
H^{-2}M_P^{2} \;,
\label{10}
\ee
with the factor of $M_P^2$ following on dimensional
grounds.
This result will  prove to be important  
a bit later on in the discussion. For now,
the reader should keep in mind that the above 
entropy is an extremely large number;
roughly, $10^{120}$ for the present-day value
of the Hubble radius.

Before moving on, we would like to make another 
pertinent observation:
Our notion of
embedding all of the exterior physics into a single
boundary surface 
 is  probably not much different 
from the principle of {\it horizon complementarity} 
\cite{banks}.~\footnote{A topical note: Horizon
complementarity can be
 viewed
as a generalization of {\it black hole
complementarity} \cite{uglum}, which
is closely related, both conceptually 
and historically, to the holographic principle.}   
The latter prescribes, for an observer 
who is (necessarily) confined to a single Hubble-sized
sphere in de Sitter space, that  any other 
Hubble sphere provides a redundant description
of the associated (quantum)  Hilbert space ---  with
this redundancy
 being 
{\it holographically} encoded
in the encompassing horizon. Although the  
nature of the surfaces are somewhat different 
(causal boundary versus de Sitter horizon), there 
is clearly a common thread with
 regard
to the storage of the exterior information.

\subsection{Uplifting}

For our argument to proceed forcefully onwards, 
it is necessary that the entanglement degrees
of freedom
be ``uplifted" to the status of 
fully physical quanta. It is not entirely clear
that such a claim could be valid, insofar 
as the entanglement in question is strictly an
observer-dependent phenomenon. To avoid any confusion,
we should point
 out that
this concern is not  at all an impediment to the
entanglement
 interpretation of
black hole entropy. In this case, the 
event horizon has a clear physical interpretation;
irrespective of the presence or not of 
observers. To put this in another (metaphorical) way,
if
a black hole falls in a vacuum, it will ``make a
sound".

In spite of this observer-dependency sticking point,
there are two  credible reasons to believe in the 
physicality of this brand of  entanglement.
Firstly, the idea of a holographic 
inducement of physical matter is not without a
precedent.
For instance, various accounts 
of the (dynamical) brane-world model \cite{kraus}
would 
lead us to believe
that some energy sources in the 
brane universe can be purely attributed to holographic
inducement;
{\it vis-a-vis}, 
the so-called {\it mirage} cosmologies
\cite{mirage}.
Of further relevance, the AdS--CFT picture of
holography
\cite{maldacena} emphatically preaches
that a quantum effect from one perspective can be
classical from
 another,
a  string can be  open and closed at the 
same time, a  regime of strong coupling can just
as easily be one of weak, the  
ultraviolet limit can be  flipped around 
into  the infrared, and so
forth.  There is certainly a lesson to be learned:  
Holographic dualities, when invoked,
can distort (and even reverse!) the 
very essence of what might have been perceived as
 objective physicality. 
So given all this, it may not be unreasonable to
expect that  
 an observer-dependent effect as 
viewed from one side of a duality  could  still be
fully physical
from the other. 

Secondly, and much less esoterically, one 
should be able to recognize a clear analogy between
the
present  consideration and  the 
Unruh effect \cite{unruh}. To review, an observer
in  (flat) Minkowski  space will, 
upon accelerating, perceive  herself as being immersed
in a bath of thermal radiation.~\footnote{Note that
the same observer
 would
measure a negative value for the 
vacuum energy (that is, negative relative to the
Minkowski vacuum),
so there is no violation of energy 
conservation going on here.} This is clearly an
observer-dependent
effect (no observer means no radiation), 
so is the perceived thermal bath real or not?
Here, we know that the answer is a 
resounding yes! If the accelerating observer happens
to
be a particle detector ---  that is, 
the famous Unruh detector --- then particles
will be registered \cite{birel-davies}, 
even though the global picture is still a trivially 
flat  spacetime.

So the moral of the story is 
that observer dependence need not be any impediment
to physicality.
 But, then again, 
should we be bothered if the dark energy (physical or
not) is
an observer-dependent phenomenon?
Here, the answer is no unless we can 
find a way to
communicate with observers that can contradict our
findings. Given the
very definition of the causal boundary, such a
communication
is, by construction, outside the realm of possibility.

With the above arguments in tow, we  feel ethically
justified in
 assuming (albeit 
tentatively~\footnote{Ultimately, 
only the ``true" theory of quantum gravity
 would
be able to dispute or assert some of our claims.
Nonetheless, for an article 
 that does advocate  the physicality
of observer-dependent phenomena (at least in
the analogous  context of acceleration horizons),
see  \cite{N-cul}. }) 
that  such an uplifting
--- from  geometric entanglement to real particles ---
can occur
and will proceed accordingly.
Now, with the uplifting taken to be  in play, it is
worth reflecting
 upon
 the status of (what we have been calling) 
the causal boundary. If these {\it holographically
induced}
particles are sufficiently energetic (as will be
verified below),
then what was once essentially  a fictitious  horizon
can become quite 
real.
That is to say, the presence of a 
dark-energy source (here, in the guise of the induced
particles) could well be sufficient to create a future
event 
horizon. [Besides the energetics, it is also necessary
that the induced
matter source has a  negative enough pressure 
({\it cf}, Footnote \ref{fn}). On this point, 
we can only speculate on precisely what breed
of particles would be suitable for this purpose.
Suffice it to say, we
 must
be talking about some highly exotic type of matter,
which is a natural 
stipulation for any discussion on what would
constitute the dark
 energy.] 

Assuming that a future horizon is indeed realized, we
now have an
 intriguing
resolution to the ``chicken-or-the-egg" conundrum that
was raised
 above:
On the one hand, the causality barrier induces an
entanglement which
 then (by
hypothesis) invokes the creation of 
real particles. On the other hand, these same
particles
uplift the status of this causal surface  to a fully
fledged
 horizon. In the sense that this  holographic surface
now has attained a
 clear
physical relevance, it adds credence to the picture of
the particles
having, themselves,  a real physical nature.  What we
have just
 described is,
in essence, a quantum-gravitational ``bootstrap"
process; whereby
the horizon and particles  are able to pull,
concurrently,   one
 another out 
from a sea of virtuality.

Finally, there is another point that  
should be emphasized. From our perspective,
the relevant infrared cutoff is (essentially) the
Hubble radius {\it
 before}
the dark energy has been induced.~\footnote{By using
the word
 ``before",
we do not mean to imply that there is 
a chronological order to this process. ``Before"
is meant, rather, in its logically precedent sense. 
Meanwhile, the relative timing
of the events in this framework will be touched upon
in the final subsection.}
Which is to say, the dark energy does
not predetermine the infrared cutoff; it is, if
anything, the other way
 around.
Hence, in our picture, the infrared cutoff can be
expected to 
asymptotically go to infinity as the energy density of
the
``regular" matter dilutes to
 nothingness.
(As an aside, such an evolution can be viewed as a 
renormalization
 group
flow~\footnote{Holographic manifestations of
a renormalization group flow are not 
uncommon. See, for instance, \cite{verlinde}.}   
 to the infrared, with the cosmic-time
coordinate  serving as the scaling parameter.)
This paints a very different picture from many other
discussions on
a holographic dark energy: Typically, the dark
energy will ultimately dominate 
the evolution of the universe and continue to do so
for the rest of time ({\it e.g.}, \cite{m-li}).
Conversely, our
 interpretation
is a dark energy that is never able to dominate over
nor be dominated
 by
the   other  sources of matter.~\footnote{An immediate
exception could be
in a very strongly gravitating 
cosmology for which the Hubble parameter is rapidly
changing. In such a regime, the Hubble radius is no
longer 
a reliable  measure of causality \cite{brustein1}.
Anyways, unless our
universe is closed, this caveat has nothing to do with
the future
 evolution.}
What is most significant about our viewpoint is the
following: The 
cosmic-coincidence  problem (why are the matter and
dark energy of the
same order in this particular era~\footnote{It is
probably
worth pointing out that, as the cosmos expand, the
dust matter dilutes  
as $a^{-3}$
whereas a true cosmological constant would be
unaffected. This is the
discrepancy that makes the ``coincidence" so
disturbing.})
would intrinsically be resolved!

One last aside: In light of the above discussion,
this horizon can not necessarily be regarded as an
event horizon {\it
 per se}.
Perhaps it should, rather, be interpreted 
as an apparent horizon that behaves like an
event horizon over
sufficiently short time scales ($t \lesssim  H^{-1}$).
Be that as it
 may,
the horizon in question would, nevertheless, be a
physically relevant one.

\subsection{Energetics}

As alluded to above, the next step 
will be to assess the energy density of these 
entanglement-inspired particles. An agreement 
with   Eq.(\ref{4}) --- that is, the originally
proposed 
(holographic) dark energy density  ---
 can be viewed as an important self-consistency
check of our framework. Conversely, a failure to do 
so  would indicate
 that 
a conceptual retooling is called for.

The ``energy of entanglement" is not as well
understood as its
entropic counterpart; however,  the very same argument
that   brought us up to Eq.(\ref{10}) can be expected
to remain
 basically
intact; namely, the entanglement energy should only
depend
on properties that are shared by the subdivided
systems. Meaning that
one would, once again, intuitively expect the
leading-order contribution
to go as the area of the common boundary surface. As
before,
rigorous calculations do bear this intuitive reasoning
out 
\cite{brustein2,brustein3}.~\footnote{One caveat
 for the energy calculation is that the expectation
value of the 
 energy
 for the {\it undivided} spacetime should be  
parametrically
smaller than the surface area of the common boundary. 
Since, in our case,  
the field theory  is to be regarded as residing in its
ground 
state [as per the parenthetic discourse just above
Eq.(\ref{10})],
 it is
 quite doubtful
that this technicality could be of any issue here.}
This deduction along with dimensional 
considerations dictates an entanglement energy 
(for the interior or exterior) of
\be
E_{E} \; \sim\; H^{-2}M_{P}^3\;.
\label{11}
\ee

 After dividing this energy by the 
interior volume ($H^{-3}$), one might suspect that
there is already
a problem with the proposal. However, 
as we will now demonstrate, this is actually not the
case. 
What has just been  
deduced is the energy as it would be measured locally
at the horizon  surface. Meanwhile,  what we  should 
really be
 examining is  
the observationally relevant energy;
that is, the energy 
as measured by the observer whose sphere of causality
defines the horizon.
Now it becomes 
especially pertinent to the discussion that the causal
boundary has,
indeed,
been elevated to a physical horizon --- as would be
expected by the
 presence of
a legitimate dark energy source. 
Having an (apparent) event horizon means that we
are really talking about an asymptotically de Sitter
spacetime.

To proceed in a quantitative fashion, 
we will require a coordinate system. For the purposes
of our ``ball-park" estimates, it will be sufficient
to  employ  a
 suitably
chosen set of de Sitter coordinates. 
Then the best choice, as relevant to an observer
(at the origin) 
with no access to the horizon exterior, is 
the well-known  static patch of de Sitter space
\cite{ds}:
\be
ds^2\;=\;-\left[1-{r^2\over H^{-2}}\right]d\tau^2\;+\;
\left[1-{r^2\over
H^{-2}}\right]^{-1}dr^2\;+\;r^2d\Omega^2_2\;.
\label{11.5}
\ee
Note that the horizon is located at $r=H^{-1}$.

What is most important, for the current evaluation, is
the
effect of 
the gravitationally induced redshift \cite{rs}.
 For an observer (at the origin) viewing any other 
point in the interior,
this effect will depend on the radial separation  and
can be quantified by the following
``redshift factor":
\be
{\cal Z}(r)\;\equiv\; \sqrt{-g_{\tau\tau}\vert_r\over
 -g_{\tau\tau}\vert_0}
\;=\; \sqrt{1-{r^2\over H^{-2}}} \;.
\label{11.75}
\ee
Near the horizon,  this factor becomes
 \be
{\cal Z}(r\lesssim H^{-1})
\;\sim\;\sqrt{{1\over H^{-1}}\left(H^{-1}-r\right)} \;
\label{12}
\ee
and, quite obviously, ${\cal Z}(r=H^{-1}) = 0\;$.

Text-book discussions on  gravitational redshifts 
tell us that 
$E_O \sim  E_{E}{\cal Z}(H^{-1})\;$,
where $E_O$ is meant to denote the energy of
observational relevance.
Naively, the vanishing redshift factor 
implies that this energy also vanishes.
However, by virtue of the  (quantum) 
uncertainty principle, it is not reasonable to presume
this degree
of localization in the calculation. A 
better rationalized approach is to calculate the
redshift 
at some ultraviolet-cutoff point. Following 
the methodology of the ``brick-wall" calculations
of black hole entropy  
\cite{thooft2} (also see \cite{frolov-buddy,myung}), 
we will judiciously place the cutoff
at a {\it proper} distance of $M^{-1}_P$ away from the
horizon. 
Recognizing that 
the proper radial distance is measured by
$dr\sqrt{g_{rr}}$
and (near the 
horizon) $g_{rr}^{-1} = -g_{\tau\tau} \sim {\cal
Z}^2\;$,
we can
then deduce that [also using
$x \equiv r/H^{-1} \lesssim  1\;$] 
\bea
M^{-1}_P \;&\sim &\; H^{-1}\int_{x}^1 {dx\over {\cal
Z}}
 \;\sim \;  H^{-1}\int_{x}^1 {dx\over \sqrt{1-x}}
\nonumber \\
& \sim & \; H^{-1}\sqrt{1-x} \;\sim\; H^{-1} {\cal
Z}\;. 
\label{13}
\eea

From the above, it  follows that
\be
{\cal Z}\;\sim\;\sqrt{1-x}\;\sim\;H M_P^{-1} \;
\label{14}
\ee
 for the near-horizon redshift and then [with the help
of
 Eq.(\ref{11})]
\be
E_O\;\sim \; {\cal Z} E_E \;\sim\; H^{-1} M_P^{2}
\label{15}
\ee
for  the observationally relevant energy. 
Dividing by the volume of the Hubble sphere,
we now obtain an energy density of
\be
\rho_{O} \;\sim\; {E_O\over H^{-3}} \;\sim\; H^2
M_P^{2} \;.
\label{16}
\ee
With the obvious identification of $H^{-1}$ with the 
infrared cutoff $L$,  this computation 
is in complete compliance with the holographic form
of Eq.(\ref{4}).\footnote{To be clear, we are not
saying here that $L$
 should
be   unequivocally identified  as  the Hubble radius
but, rather, that these
 length
scales will be of the same order of magnitude. That is
to say, 
we would not anticipate
any conflict with observation  in the sense suggested
by \cite{hsu}.}

\subsection{Non-Locality}

One might be bothered by a dark energy that is
strongly localized at a
 horizon.
Indeed, such an energy source could 
very well be detectably different from the
cosmological constant
that it  aspires to replicate. In our picture, the 
horizon  certainly represents the
surface  of inducement (via entanglement), but a
simple argument  
will reveal  that the dark energy must necessarily
delocalize and
fill up the Hubble sphere.

Let us now explicate this point. To discuss the
locality (or lack
 thereof)
of a given particle, it is usually sufficient to
consider its Compton
wavelength $\lambda_C\;$.
By recalling the origin of the dark energy
as being a process of entanglement, we can immediately
constrain the
 associated
 $\lambda_C$ from above. Significantly, the 
entanglement procedure entailed tracing over
{\it all} of the degrees of freedom to the exterior of
(what is now) the horizon.
 Hence,
insofar as it  ``concerns" a dark-energetic particle,
the universe
must end at the horizon; which is to say, 
there is, practically speaking, no longer an
exterior region to speak of.
And so, it must follow that 
\be
\lambda_C \;\leq \; H^{-1} \;.
\label{17}
\ee 
 
To constrain $\lambda_C$ from below, we need only
to invoke the usual statistical interpretation of an
entropy,
along with the uncertainty principle.  
Applying the former input to Eq.(\ref{10}),
we count  the number of particles 
to be  (roughly) $N \sim  H^{-2}M^{2}_{P}\;$;
meaning that
the energy per particle is 
$E_O/N \sim  H\;$.  As dictated by the uncertainty
principle, the inverse of this ratio provides a lower
bound for
the spatial extent of any given particle, or
\be
\lambda_C \;\geq \; H^{-1} \;.
\label{18}
\ee 
 
There is, obviously, only one
way to 
 avoid a contradiction between Eqs.(\ref{17}) and
(\ref{18});
namely,
\be
\lambda_C \; \sim \; H^{-1} \;.
\label{19}
\ee 
 And so, not only are the particles not localized at
the horizon, they
 are
delocalized across the full extent of the Hubble
sphere!

The picture is now the following: We started with
degrees of freedom
 being excited  on 
a spherical surface, and then end up with a
sphere-filling gas of
 extremely
long-wavelength and inert~\footnote{We say ``inert"
because
these particles should, by default, be able to mimic a
cosmological
 constant.
At this point, the property of  
inertness must be put in by hand; however, see below.}
 particles.
Note that there is no need to establish a {\it
dynamical} mechanism
for this delocalization to happen. From a holographic
vantage point,
the cosmic-time coordinate 
is (as mentioned above) the scale factor of a
renormalization
group flow: Meaning that the flow of cosmic time
coincides with
the changes in the scale of the infrared cutoff ($L
\sim  H^{-1}$) but
 has
nothing to say about the  occurrence of
holographically
triggered mechanisms. These take place in some
abstract
(from our perspective) realm which is governed by
a time evolution that is --- in all likelihood ---
unrelated
to the flow of cosmic time. And so, cosmically
speaking, the particles
have always been there and are always delocalized in
accordance
with the scale set, at any given time, by $L =
L(t)$.~\footnote{Let
us point out, once again, that $L$ might be much
different
than $H^{-1}$ when the universe is strongly
gravitating;
in particular, in the earliest stages of the
cosmological
evolution.}

Nevertheless, one would
still be entitled to ask if there is a physical
framework
that supports such a dramatic degree  of ubiquity.
That is to say, is there any credible  reason to
believe
in the existence of particles with such an unnaturally
long wavelength? We will now  proceed to argue that
this type of delocalization is actually
quite natural in the current context.

To begin here,  let us ponder as to what might be the
nature of
these holographically induced particles. The key
word here is holograhic(ally): The pertaining
principle
is often asserted to be a statement about --- or even
a manifestation
of --- quantum gravity. Alternatively, regarding this
principle 
as an effect rather than a cause,
one might prefer to
focus on the entanglement origin of these particles
and just leave holography out of it.
But, even then, 
entanglement entropy  can similarly be viewed as a
synthesis of 
 gravitational and  quantum concepts;
inasmuch as geometry (which 
underlies entanglement) is the  very essence of
gravity.
So, by all accounts, these particles
must, on some level, be
identifiable as ``quanta of geometry".~\footnote{We
have
purposely  avoided saying
 ``quanta of gravity", which might be misconstrued 
as a synonym for gravitons.}  

Continuing along this line of inquiry, what (if
anything) can we say
 about
such geometric quanta? 
Undoubtedly, the answer will be highly model
dependent.
However, it is notable that both string theory
(especially in its
 matrix
realization \cite{banks2}) and loop quantum gravity
\cite{loop} share 
something in common when informing us about
their  fundamental quanta; respectively,  D0-branes
and spin junctions.
For {\it both} of these models, the 
fundamental constituents  are expected to  behave,
collectively, 
as a gas of {\it non-interacting} and
 fully {\it distinguishable} particles 
\cite{strominger,volovich,banks3,minic1,krasnov,rovelli,setter}.
This is a much different state of affairs from the
``garden-variety"
particles ---
bosons and fermions --- which are (of course) 
always treated as being indistinguishable.

What is  particularly interesting (for us anyways)
about 
distinguishable particles is that they obey a
generalized
form of Bose and Fermi statistics that is known
as {\it infinite statistics} \cite{greenberg}. 
More formally,
\be
a_ja_k^{\dagger} \;-\; qa_k^{\dagger}a_j
\;=\;\delta_{jk}\;,
\label{20}
\ee
where $a_j$  and $\;a^{\dagger}_k\;$ represent the
usual lowering
and raising operators, while
$q$ is a real-number parameter that can range between
$-1$
(the fermionic limiting case) and $+1$ (the bosonic
limit). 
Thanks to the above symbolic convention,
such particles are often referred to as {\it quons}.

Now we finally come to the punch line. Infinite
statistics
will  inevitably describe a theory that 
is inherently non-local.  To
see this, it is sufficient to consider the number
operator; which can be suitably defined by the
commutator relation
\be
\left[N_j,a_k^{\dagger}\right]
\;=\;\delta_{jk}a_k^{\dagger}\;.
\label{21}
\ee
Choosing the $q=0$ case for simplicity, one can
readily
be convinced  that the number operator must take on
the
recursive form
\be
N_j\;=\;a^{\dagger}_ja_j \;+\;\sum_{k}a^{\dagger}_k
a^{\dagger}_ja_ja_k \;+\;
\sum_{l}\sum_{k}a^{\dagger}_l
a^{\dagger}_ka^{\dagger}_ja_ja_ka_l \;+\; ... \;.
\label{22}
\ee
And so one finds that, even for the simplest  ($q=0$)
case, the resulting theory is a non-local one,
and the expansions are even more convoluted when 
$0 < \vert q\vert < 1\;$.~\footnote{It is worth
pointing out
that this non-locality does not jeopardize some
of the more desirable features of conventional quantum
theory such as
the TCP theorem and cluster decomposition. So if
non-locality
is, in itself, not an issue, the quantum practitioner
need not be overly concerned about  harmful ``side
effects".}

Let us tie this altogether. Our dark energy particles
must be delocalized in accordance with Eq.(\ref{19}).
Meanwhile, whatever these particles exactly are
(dubbed as
quanta of geometry above), the consensus view 
of quantum-gravity theorists is that they should be 
fully distinguishable
and, thereby, obey infinite statistics. Conveniently,
infinite statistics can describe a  non-local yet
otherwise sensible
quantum field
theory. Meaning that, given the quantum-gravitational
pedigree of the particles, this delocalization of
the dark energy turns out to be quite natural (and
perhaps
even  to be anticipated).
As an added bonus, the fundamental particles of
quantum
gravity  are expected to be
 non-interacting, which is an almost essential
stipulation for  the (presumably) inert substance
that  would constitute the dark energy.

Before concluding, we would be remiss
 to not mention a pair of recent papers
that have already made a connection between the dark
energy
and infinite statistics \cite{ng,minic2}. These
studies certainly
inspired a substantial amount of the current
subsection, and we  hope
that our perspective was sufficiently different
to contribute to the discussion.

\section{Overview}

The holographic model of dark energy \cite{m-li}
(including
the many variations thereof \cite{hde-rev}) has proven
to be
a promising avenue for understanding the
current acceleration of the universe.
As already reflected upon, this model
has been  demonstrated to have  
 an ample amount of  
potential for matching the
empirically based expectations of  the
observable universe; with regard to both the
past and the present. What is still lacking, however,
is a viable explanation of what the holographic dark
energy is exactly supposed to represent. That is to
say,
what is the composition of the implicated energy
source  and why does 
this substance  exhibit the properties that are
desirable
for mimicking a cosmological constant?

In the current paper, we have made an initial
attempt
at filling  in this gap. Many of our arguments
were admittedly speculative and many details
were regrettably left out. Nonetheless, we still feel
that
some modest  progress has  been achieved along the
stated lines.
Our construction follows a logical succession
of ideas with a self-consistent prediction
for the energy density.
Moreover, just by thinking about the issues
``out loud", we have been able to make some deductions
that have been previously missed or, at least,
mostly overlooked. Most significantly, our
discussion has cast a new light upon the relationship
between the holographic dark energy and its associated
infrared cutoff. From our perspective, the latter
is an antecedent for the former; meaning that
the magnitude of the dark energy is
fixed to (more or less) match the other
matter sources in the universe. If this
turns out to be a valid assessment of
the situation, then the (so-called) cosmic-coincidence
problem would be quite naturally resolved.

Obviously, there is much more work to be done;
either in addressing the outstanding questions of the
current
proposal or in constructing an altogether different
framework that starts anew. It is hoped that
both paths are enthusiastically followed; not only by
the current
author but by others in the field as well.

\newpage
\section*{Acknowledgments}
Research is financially supported by the University of
Seoul.
The author thanks Yun Soo Myung for
his inputs and valued discussions, and CQUeST  at
Sogang University
for their hospitality. The author also thanks Stephen
Hsu for
pointing out an oversight in the first rendition of
the
manuscript.



\begin{thebibliography}{99}

\bibitem{acc-uni} S. Perlmutter {\it et al}., Nature
{\bf 391}, 51
 (1998).
\bibitem{ein-cc} 
A. Einstein, {\it Sitzungsber. Preuss. Akad. Wiss}.
phys.-math. Klasse
 VI, 142
(1917).
\bibitem{weinberg} S. Weinberg, Rev. Mod. Phys. {\bf
61}, 1 (1989);\\
 ``The Cosmological Constant Problems",
arXiv:astro-ph/0005265 (2000).
\bibitem{cc-rev1}  S.M. Carroll,    Living Rev. Rel.
{\bf 4}, 1 (2001)
  [arXiv:astro-ph/0004075].
\bibitem{cc-rev2} P.J.E. Peebles 
and  B. Ratra, Rev. Mod. Phys. {\bf 75}, 559 (2003) 
 [arXiv:astro-ph/0207347]. 
\bibitem{cc-rev3} 
T. Padmanabhan, ``Dark Energy 
and Gravity",  arXiv:0705.2533 and to appear in Gen.
Rel. Grav. (2007).
  \bibitem{quin}
C. Wetterich, Nucl. Phys. B {\bf 302}, 668 (1988).
\bibitem{infl} A.A. Starobinsky, Phys. Lett. B {\bf
91}, 99 (1980).
\bibitem{infl2} A. Guth, Phys. Rev. D {\bf 23}, 347
(1981).
\bibitem{m-li}  M. Li, 
Phys. Lett. B {\bf 603}, 1 (2004)
[arXiv:hep-th/0403127].
\bibitem{N-myung} Y.S. Myung, Phys. Lett. B {\bf 610},
18 (2005) 
[arXiv:hep-th/0412224].
\bibitem{hde-rev} M. Li, C. Lin 
and Y. Wang, ``Some Issues Concerning Holographic Dark
Energy", 
arXiv:0801.1407 (2008).
\bibitem{bekenstein} J.D. Bekenstein, Phys. Rev. D
{\bf 23}, 287
 (1981).
\bibitem{thooft} G. 't Hooft, ``Dimensional 
Reduction in Quantum Gravity", arXiv:gr-qc/9310026
(1993).
\bibitem{susskind} L. Susskind, J. Math. Phys. 
{\bf 36}, 6377 (1995) [arXiv:hep-th/9409089].
\bibitem{bousso} R. Bousso, Rev. 
Mod. Phys. {\bf 74}, 825 (2002)
[arXiv:hep-th/0203101].
\bibitem{cohen} A.G. Cohen, D.B. Kaplan 
and A.E. Nelson, Phys. Rev. Lett. {\bf 82}, 4971
(1999) 
[arXiv:hep-th/9803132].
\bibitem{area}  J.D. Bekenstein, Lett. Nuovo. Cim.
{\bf 4}, 737
 (1972);\\
 Phys. Rev. D {\bf 7}, 2333 (1973); Phys. Rev. D {\bf
9}, 3292 (1974).
\bibitem{hoop}  K. S. Thorne, 
in {\it Magic without Magic: John Archibald Wheeler},
ed. J. Klauder
 (Freeman,
San Francisco, 1972).
\bibitem{non-satur} D. Pavon 
and W. Zimdahl, Phys. Lett. B {\bf 628}, 206 (2005) 
[arXiv:gr-qc/0505020]. 
\bibitem{N-od} S. Nojiri and S.D. Odintsov,
Gen. Rel. Grav. {\bf 38}, 1285 (2006)
[arXiv:hep-th/0506212]. 
\bibitem{cai} R.-G. Cai, Phys. Lett B {\bf 657}  228
(2007)
[arXiv:0707.4049]. 
\bibitem{hsu} S.D.H. Hsu, Phys. 
Lett. B {\bf 594}, 13 (2004) [arXiv:hep-th/0403052].
\bibitem{m-li2} Q.-G. Huang and  M. Li,  JCAP {\bf
0408},  013 (2004) 
[arXiv:astro-ph/0404229].
\bibitem{time-grav} R. Horvat, Phys. Rev. D {\bf 70},
087301 (2004) 
 [arXiv:astro-ph/0404204].
\bibitem{interact}  W. Zimdahl 
and D. Pavon, ``Interacting holographic dark energy", 
 arXiv:astro-ph/0606555 (2006).
\bibitem{N-rand} B. Guberina, R. Horvat and H. Nikolic,
 Phys. Rev. D {\bf 72}  125011 (2005) 
[arXiv:astro-ph/0507666]. 
\bibitem{bh-ent1} L. Bombelli, R. Koul, J. Lee 
and R. Sorkin, Phys. Rev. D {\bf 34}, 373 (1986).
\bibitem{bh-ent2} M. Srednicki, Phys. Rev. D {\bf 71},
66 (1993).
\bibitem{ads-ent1} S. Ryu 
and T. Takayanagi, Phys. Rev. Lett. {\bf 96}, 181602
(2006) 
[arXiv:hep-th/0603001]. 
\bibitem{ads-ent2} D.V. Fursaev, JHEP {\bf 0609}, 018
(2006)
[arXiv:hep-th/0606184]. 
\bibitem{maldacena} J.M. Maldacena, Adv. Theor. Math.
Phys.
{\bf 2}, 231 (1998) [arXiv:hep-th/9711200].
\bibitem{ent-de}  J.-W. Lee, J. Lee 
and H.-C. Kim, JCAP {\bf 08},  005 (2007)   
[arXiv:hep-th/0701199]. 
\bibitem{ent-de2}  J.-W. Lee, J. Lee 
and H.-C. Kim,
``Quantum Informational Dark Energy: Dark energy from
forgetting", 
arXiv:0709.0047 (2007). 
\bibitem{horvat} R. Horvat, ``Holographic 
dark energy: quantum correlations against
thermodynamical description",
 arXiv:0711.4013 (2007).
\bibitem{N-greek} E.N. Saridakis,
``Restoring Holographic Dark Energy in Brane
Cosmology", arXiv:0712.2228
and to appear in Phys. Lett. B (2007). 
\bibitem{unruh} W.G. Unruh,
Phys. Rev. D {\bf 14}, 870 (1976). 
\bibitem{brustein2} R. Brustein, 
D. Eichler, S. Foffa and D.H. Oaknin, Phys. Rev. D
{\bf65}, 105013
(2002)  [arXiv:hep-th/0009063]. 
\bibitem{brustein3} R. Brustein and A. Yarom, JHEP
{\bf 0501}, 046
 (2005)
[arXiv:hep-th/0302186]. 
\bibitem{brustein1} R. Brustein 
and G. Veneziano, Phys. Rev. Lett. {\bf 84}, 5695
(2000) 
[arXiv:hep-th/9912055].
\bibitem{epr}  A. Einstein, B. Podolsky and N. Rosen,
Phys. Rev. {\bf
 47},
777 (1935).
\bibitem{adami} N.J. Cerf and C. Adami,  ``Quantum
Mechanics of
 Measurement",
 arXiv:quant-ph/9605002 (1996).
\bibitem{N-hsu} R. Buniy and S. Hsu, Phys. 
Lett. B {\bf 644}, 72 (2007) 
[arXiv:hep-th/0510021].
\bibitem{N-req} M. Requardt, ``Entanglement-Entropy
for 
Groundstates, Low-lying and Highly Excited Eigenstates
of General (Lattice) Hamiltonians",
arXiv:hep-th/0605142 (2006).
\bibitem{banks} T. Banks 
and W. Fischler, ``M-Theory Observables for
Cosmological Spacetimes",
arXiv:hep-th/0102077 (2001).
\bibitem{uglum} L. Susskind, L. Thorlacius and J.
Uglum, 
Phys. Rev. D {\bf 48}, 3743 (1993)
[arXiv:hep-th/9306069].
\bibitem{kraus} P. Kraus, JHEP {\bf 9912}, 011 (1999)
[arXiv:hep-th/9910149]. 
\bibitem{mirage}  A. Kehagias and E. Kiritsis, JHEP
{\bf 9911}, 022
 (1999)  
[arXiv:hep-th/9910174].  
\bibitem{birel-davies} N.D. Birrell  and P.C.W.
Davies, {\it Quantum
 fields in
curved space} (Cambridge University Press, Cambridge,
1982).
\bibitem{N-cul} H. Culetu, Int. J. Mod. Phys. D {\bf
15}, 2177 (2006) 
[arXiv:hep-th/0607049].
\bibitem{verlinde} J. de Boer, E. Verlinde 
and H. Verlinde, JHEP {\bf 0008}, 003 (2000)
 [arXiv:hep-th/9912012]. 
\bibitem{ds} M. Spradlin, A. Strominger and A.
Volovich, 
``Les Houches Lectures on De Sitter Space", 
arXiv:hep-th/0110007 (2001).
\bibitem{rs} R. Tolman and P. Ehrenfest, Phys. Rev.
{\bf 36}, 1761
 (1930).
\bibitem{thooft2} G. 't Hooft, Nucl. Phys. B {\bf
256}, 727 (1985).
\bibitem{frolov-buddy} S. Mukohyama, M. Seriu 
and H. Kodama, Phys. Rev. D {\bf 55}, 7666 (1997)
[arXiv:gr-qc/9701059].
\bibitem{myung} Y.S. Myung, Phys. Lett. B {\bf 636},
324 (2006)  
[arXiv:gr-qc/0511104]. 
\bibitem{banks2} T. Banks, W. Fischler, S.H. Shenker
and L. Susskind, 
Phys. Rev. D {\bf 55}, 5112 (1997)  
[arXiv:hep-th/9610043]. 
\bibitem{loop} A. Ashtekar, C. Rovelli 
and L. Smolin,  Phys. Rev. Lett. 69, 237 (1992) 
[arXiv:hep-th/9203079].
\bibitem{strominger}  A. Strominger, Phys. Rev. Lett.
{\bf 71}, 3397
  (1993) 
[arXiv:hep-th/9307059].
\bibitem{volovich} I.V. Volovich ``D-branes, 
Black Holes and $SU(\infty)$ Gauge Theory", 
arXiv:hep-th/9608137   (1996). 
\bibitem{banks3} T. Banks, W. Fischler, I.R. Klebanov 
and L. Susskind, JHEP {\bf 9801}, 008 (1998)  
[arXiv:hep-th/9711005]. 
\bibitem{minic1} D. Minic, ``Infinite Statistics 
and Black Holes in Matrix Theory", 
arXiv:hep-th/9712202  (1997).
 \bibitem{krasnov} K.V. Krasnov,  Phys. Rev. D {\bf
55}, 3505 (1997) 
[arXiv:gr-qc/9603025]. 
\bibitem{rovelli} C. Rovelli,  Phys. Rev. Lett. {\bf
77}, 3288 (1996) 
[arXiv:gr-qc/9603063]. 
\bibitem{setter} S.A. Major 
and K.L. Setter, Class. Quant. Grav. 18, 5125 (2001)  
[arXiv:gr-qc/0101031].
\bibitem{greenberg} O.W. Greenberg, ``Quons, 
an interpolation between bose and fermi oscillators", 
arXiv:cond-mat/9301002 (1993). 
\bibitem{ng}  Y.J. Ng, ``Holographic Foam, 
Dark Energy and Infinite Statistics", 
arXiv:gr-qc/0703096 and to appear in Phys. Lett. B
(2007). 
\bibitem{minic2} V. Jejjala, M. Kavic 
and D. Minic, ``Fine Structure of Dark Energy and New
Physics",  
arXiv:0705.4581 (2007).




\end{thebibliography}
\end{document}